\documentclass{modified.ws-p10x7}

\begin{document}

\title{STATUS OF THEORETICAL $\bar{B} \to X_s \gamma$ AND $\bar{B} \to X_s l^+ l^-$ ANALYSES}

\author{MIKO{\L}AJ MISIAK}

\address{Theory Division, CERN, CH-1211 Geneva 23, Switzerland}

\twocolumn[\maketitle\abstract{ Status of the theoretical $\bar{B} \to
  X_s \gamma$ and $\bar{B} \to X_s l^+ l^-$ analyses is reviewed.
  Recently completed perturbative calculations are mentioned.
The level at which non-perturbative effects are controlled is discussed.}]

The present talk will be devoted to discussion of the SM predictions
only. Let us begin with $\bar{B} \to X_s \gamma$.  Since the
completion of NLO QCD calculations\cite{NLO} 4 years ago, many new
analyses have been performed. They include evaluation of
non-perturbative $\Lambda^2/m_c^2$ corrections\cite{mc} and the
leading electroweak corrections.\cite{ew,KN99} None of these results
exceeds half of the overall $\sim \hspace{-1mm} 10\%$ uncertainty, and
there are cancellations among them. In consequence, the prediction for
$BR[\bar{B} \to X_s \gamma]$ remains almost unchanged: $(3.29 \pm
0.33) \times 10^{-4}$. This prediction agrees very well with the
measurements of CLEO\cite{CLEO99a}, ALEPH\cite{ALEPH98} and
BELLE\cite{BELLE00}, whose combined result is $(3.21 \pm 0.40) \times
10^{-4}$.

The dominant contribution to the perturbative $b \to s \gamma$
amplitude originates from charm-quark loops. After including QCD
corrections, the top-quark contribution is less than half of the
charm-quark one, and it comes with an opposite sign. This fact should
be remembered when one attempts to extract $|V_{ts}|$ from $b \to s
\gamma$. The $u$-quark contribution is suppressed with respect to the
charm one by $|V_{ub} V_{us}|/|V_{cb} V_{cs}| \simeq 2\%$.

The results of CLEO, ALEPH and BELLE have to be understood as the ones
with subtracted intermediate $\psi$ background, i.e. the background
from $\bar{B} \to \psi X_s$ followed by $\psi \to X' \gamma$. This
background gives more than $4 \times 10^{-4}$ in the ``total'' BR, but
gets suppressed when only high-energy photons are counted. A rough
estimate\cite{KM00} of the effect of the photon energy cutoff on this
background can be made when $X_s$ in $\bar{B} \to \psi X_s$ is assumed
to be massless, and the non-zero spin of $\psi$ is ignored.
Then,\footnote{
  The $\psi \to X \gamma$ spectrum is available from the ancient
  MARK~II data\cite{MARK2}. New results are expected soon
  from the BES experiment in Beijing.}
the intermediate $\psi$ background is less than 5\%, for the present
experimental cutoff $E_{\gamma} > 2.1$~GeV in the $\bar{B}$-meson rest
frame.\footnote{
  A further suppression (to less than 1\%) is found when $X_s$ is not
  treated as massless but the measured\cite{CLEO95} mass spectrum is
  used.}
However, the background grows fast when the cutoff goes down.  

The photon energy cutoff {\em will} have to go down by at least 200 or
300 MeV in the future. With the present one, non-perturbative effects
related to the unknown $\bar{B}$-meson shape function\cite{KN99}
considerably weaken the power of $b \to s \gamma$ for testing new
physics. For the same reason, future measurements of $\bar{B} \to X_s
\gamma$ should rely as little as possible on theoretical predictions
for the precise shape of the photon spectrum above $E_{\gamma} \sim
2$~GeV.  

A systematic analysis of non-\linebreak -perturbative effects in
$\bar{B} \to X_s \gamma$ at order ${\cal O}(\alpha_s(m_b))$ is
missing. There is no straightforward method to perform such an
analysis, because there is no obvious operator product expansion for
the matrix elements of the 4-quark operators, in the presence of one
or more hard gluons (i.e. the gluons with momenta of order $m_b$). At
present, we have only intuitive arguments to convince ourselves that
such non-perturbative effects are probably significantly smaller than
the overall $\sim \hspace{-1mm} 10\%$ theoretical uncertainty in
$BR[\bar{B} \to X_s \gamma]$, when the energy cutoff is between 1 and
2 GeV, and when the intermediate $\psi^{(}$$'$$^{)}$ contribution(s)
are subtracted.

As far as the decay $\bar{B} \to X_s l^+ l^-$ is concerned (for $l=e$
or $\mu$), the best control over non-perturbative effects can be
achieved in the region of low dilepton invariant mass ($\hat{s} \equiv
m^2_{l^+ l^-}/m_b^2 \in [0.05,\;0.25]$).  The present
prediction\cite{CMU00} for the branching ratio integrated over this
domain is $(1.46 \pm 0.19) \times 10^{-6}$. The quoted uncertainty is
only the perturbative one. The non-perturbative $\Lambda^2/m_c^2$ and
$\Lambda^2/m_b^2$ contributions\cite{HQee} have been included in the
central value. They are around 2\% and 5\%, respectively.

A calculation of ${\cal O}(\alpha_s)$ terms in all the relevant Wilson
coefficients $C_i(m_b)$ has been recently completed\cite{CMU00}, up to
small effects originating from 3-loop RGE evolution of $C_9$.
However, the perturbative uncertainty in the above-mentioned
prediction remains close to $\sim \hspace{-1mm} 13\%$, because 2-loop
matrix elements of the 4-quark operators are unknown.

The low-$\hat{s}$ branching ratio is as sensitive to new physics as
the forward-backward or energy asymmetries, i.e. $\sim \hspace{-1mm}
100\%$ effects are observed when $C_7(m_b)$ changes sign.

The background from $\bar{B} \to \psi X_s$ followed by $\psi \to l^+
l^-$ is removed by the cutoff $\hat{s} < 0.25$. Analogous
contributions from virtual $c\bar{c}$ states are, in principle,
included in the calculated $\Lambda^2/m_c^2$ correction. An
independent verification of this fact can be performed with help of
dispersion relations and the factorization approximation.\cite{KS96}
Indeed, for $\hat{s} < 0.25$, the difference between results obtained
with help of the two methods is quite small, and can be attributed to
higher-order perturbative effects.

On the other hand, the background from $\bar{B} \to \psi X_s$ followed
by $\psi \to X' l^+ l^-$ has never been studied. Most probably, for
$\hat{s} < 0.25$, it is less important than the analogous background
in the case of $\bar{B} \to X_s \gamma$. Experiment-based calculations
of these backgrounds are awaited, because they are essential for
performing theoretical estimates of similar non-perturbative
contributions from other $c\bar{c}$ states.

\newcommand{\np}[3]{ {\it Nucl. Phys.} B {\bf #1}, #3 (#2)}
\newcommand{\pl}[3]{ {\it Phys. Lett.} B {\bf #1}, #3 (#2)}
\newcommand{\pr}[3]{ {\it Phys. Rev.} D {\bf #1}, #3 (#2)}
\newcommand{\prl}[3]{ {\it Phys. Rev. Lett.} {\bf #1}, #3 (#2)}

\end{document}